\documentstyle[12pt]{article}
\advance\voffset by -0.8in
\advance\hoffset by -0.55in
\begin{document}
\newcommand{\reg}[1]{(\ref{#1})}
\newcommand{\sqr}[2]{{{\vcenter{\vbox{\hrule height.#2pt
\hbox{\vrule width.#2pt height#1pt \kern#1pt
\vrule width.#2pt}
\hrule height.#2pt}}}}}
\newcommand{\square}{{\mathchoice\sqr{17}4\sqr{17}4\sqr{13}3\sqr{13}3}}
\newcommand{\Hil}{{\cal H}}
\newcommand{\C}{{\cal C}}
\newcommand{\ze}{{\bf Z}}
\title{Discussion of Self-Dual $c=24$ Conformal Field Theories}
\author{P.S. Montague\\
Department of Applied Mathematics and Theoretical Physics\\
University of Cambridge\\
Silver Street\\
Cambridge CB3 9EW\\
U.K.}
\maketitle
\begin{abstract}
We discuss questions arising from the recent work of
Schellekens\cite{Schell:Venkov},
and also from an earlier paper by Schellekens and
Yankielowicz\cite{SchellYank:curious}.
We summarise Schellekens' results, and proceed to discuss the uniqueness
of the $c=24$ self-dual conformal field theory with no weight one
states, {\em i.e.} the Monster module $V^\natural$\cite{FLMproc}. After
introducing
the concept of complementary representations, we examine $\ze_2$-orbifold
constructions in general, and then proceed to apply our
considerations firstly to the specific case of the FKS constructions
$\Hil(\Lambda)$ and then to the reflection twisted theories
$\widetilde\Hil(\Lambda)$\cite{DGMtwisted}. Our techniques provide evidence for
the existence of several new theories beyond those proven to exist in
\cite{DGMtwisted} and conjectured to exist in \cite{SchellYank:curious}.
\end{abstract}
\section{Introduction}
Recently much progress has been made
towards the classification of CFT's. One approach to this problem is to study
the algebra of the fusion rules of representations of some chiral algebra (an
extension of the Virasoro algebra)\cite{Verlinde}. However, we are
interested here in
CFT's which are overlooked by this technique, {\it i.e.} theories whose fusion
rules are trivial, though they themselves are not necessarily without an
interesting structure. (Indeed, one such theory is the
natural module $V^\natural$ for the Monster,
first constructed by Frenkel, Lepowsky and
Meurman\cite{FLMbook,FLMacadsci,FLMproc}.)
Such theories clearly must be classified separately
>from the mainstream if fusion rule techniques are to be used.

We consider chiral bosonic meromorphic CFT's defined on the Riemann sphere (see
\cite{PGmer,thesis} for the relevant definitions, and also \cite{FHL}
for a mathematician's view). We define a CFT $\Hil$ to be self-dual if the
partition function
\begin{equation}
\chi_\Hil(\tau)={\rm Tr}_\Hil\,q^{L_0-c/24}\,,
\end{equation}
where $q=e^{2\pi i\tau}$, is covariant under modular transformations
of $\tau$, {\em i.e.} invariant under $S:\tau\mapsto-1/\tau$ and
invariant up to a phase under $T:\tau
\mapsto\tau+1$. (So that the full partition function when the
antichiral sector is included is then modular invariant, as required
for the theory to be physically well-defined on the torus described by
the parameter $\tau$.) This restricts us to $c\in 8\ze$. The theories
for $c=8$ and $c=16$ are easily classified\cite{PGmer}. They are
simply the FKS constructions $\Hil(\Lambda)$ from the even self-dual
lattices in the corresponding dimensions, {\em i.e.} the root lattice
of $E_8$ in 8 dimensions and the lattices ${E_8}^2$ and ${D_{16}}^+$
(an extension of the root lattice of $D_{16}$ by adding in one of the
spinor weights) in 16 dimensions.

Only in 24 dimensions does the problem of classification first become
non-trivial. Indeed, it may be argued that, since the classification
of even the even self-dual lattices in more than 24 dimensions is
intractable at present due to the rapid increase in their number, then
$c=24$ is really the only case worthy of consideration. There are 24
inequivalent even self-dual lattices in 24 dimensions\cite{ConSlo}.
The constructions
$\Hil(\Lambda)$ and $\widetilde\Hil(\Lambda)$ of \cite{DGMtwisted} would thus
be naively expected to produce 48 CFT's. However, it is shown in
\cite{DGMtrialsumm}
and \cite{DGMtriality} that the constructions produce equivalent theories if
and only if there is a corresponding doubly-even self-dual binary code. There
are
9 such codes in 24 dimensions\cite{binary:class}, and so we obtain 39 distinct
self-dual $c=24$ CFT's. These are, so far, the only such theories which have
been constructed explicitly, though in \cite{SchellYank:curious} two further
theories were postulated to exist. We shall discuss these further in the
following
section.
\section{Schellekens' results}
\label{Schell}
In \cite{Schell:Venkov} Schellekens proves results for conformal field
theories analogous to those
obtained by Venkov in his reformulation of Niemeier's classification of even
self-dual lattices\cite{Venkov,ConSlo}.
In particular, it is shown that the Kac-Moody
algebra generated by the modes of the weight one states\cite{PGmer}
is restricted to contain components whose central charges sum to 24
and moreover have a common value for the ratio of the Coxeter number to the
level, given in terms of the number $N$ of weight one states by
\begin{equation}
{g\over k}={N\over{24}}-1\,.
\end{equation}
The possible combinations of algebras thus allowed in a $c=24$ self-dual
conformal field theory with $N$
weight one states are shown below. The rank is also indicated
for convenient reference. A $*$ indicates that the theory is one of the 39
obtained by the FKS construction or a reflection twist of such a
theory.
The two theories marked by a $\dagger$ are those claimed (but not proven)
to exist in \cite{SchellYank:curious}. Their evidence consisted of the
construction
of a modular invariant combination of characters of the corresponding Kac-Moody
algebras, though no explicit constructions were given. We shall return to this
question
in a later section.
We mark by $\oplus$ new theories proposed/constructed in this work.
\begin{tabbing}
$N\hskip30pt$  \= algebra$\hskip70pt$ \= rank\hskip50pt \=
$N\hskip30pt$  \= algebra$\hskip70pt$ \= rank \\
$0$ \> $\emptyset$ \> 0 * \>
$24$ \> $U(1)^{24}$ \> 24 * \\
$25$ \> ${A_{3,96}}{C_{2,72}}$ \> 5 \>
$25$ \> ${A_{1,48}}{A_{2,72}}{G_{2,96}}$ \> 5 \\
$25$ \> ${A_{1,48}}^{3}{A_{2,72}}^{2}$ \> 7 \>
$25$ \> ${A_{1,48}}^{5}{C_{2,72}}$ \> 7 \\
$26$ \> ${A_{2,36}}^{2}{C_{2,36}}$ \> 6 \>
$26$ \> ${A_{1,24}}{A_{2,36}}{A_{3,48}}$ \> 6 \\
$26$ \> ${A_{1,24}}^{2}{C_{2,36}}^{2}$ \> 6 \>
$26$ \> ${A_{1,24}}^{4}{G_{2,48}}$ \> 6 \\
$26$ \> ${A_{1,24}}^{6}{A_{2,36}}$ \> 8 \>
$27$ \> ${A_{1,16}}{C_{2,24}}{G_{2,32}}$ \> 5 \\
$27$ \> ${A_{1,16}}{A_{4,40}}$ \> 5 \>
$27$ \> ${A_{1,16}}{A_{2,24}}^{3}$ \> 7 \\
$27$ \> ${A_{1,16}}^{2}{C_{3,32}}$ \> 5 \>
$27$ \> ${A_{1,16}}^{2}{B_{3,40}}$ \> 5 \\
$27$ \> ${A_{1,16}}^{3}{A_{2,24}}{C_{2,24}}$ \> 7 \>
$27$ \> ${A_{1,16}}^{4}{A_{3,32}}$ \> 7 \\
$27$ \> ${A_{1,16}}^{9}$ \> 9 \>
$28$ \> ${G_{2,24}}^{2}$ \> 4 \\
$28$ \> ${D_{4,36}}$ \> 4 \>
$28$ \> ${A_{2,18}}{C_{2,18}}^{2}$ \> 6 \\
$28$ \> ${A_{1,12}}{A_{3,24}}{C_{2,18}}$ \> 6 \>
$28$ \> ${A_{1,12}}^{2}{A_{2,18}}{G_{2,24}}$ \> 6 \\
$28$ \> ${A_{1,12}}^{4}{A_{2,18}}^{2}$ \> 8 \>
$28$ \> ${A_{1,12}}^{6}{C_{2,18}}$ \> 8 \\
$30$ \> ${C_{2,12}}^{3}$ \> 6 $\oplus$ \>
$30$ \> ${A_{3,16}}^{2}$ \> 6 \\
$30$ \> ${A_{2,12}}^{2}{G_{2,16}}$ \> 6 \>
$30$ \> ${A_{1,8}}^{2}{C_{2,12}}{G_{2,16}}$ \> 6 \\
$30$ \> ${A_{1,8}}^{2}{A_{4,20}}$ \> 6 \>
$30$ \> ${A_{1,8}}^{2}{A_{2,12}}^{3}$ \> 8 \\
$30$ \> ${A_{1,8}}^{3}{C_{3,16}}$ \> 6 \>
$30$ \> ${A_{1,8}}^{3}{B_{3,20}}$ \> 6 \\
$30$ \> ${A_{1,8}}^{4}{A_{2,12}}{C_{2,12}}$ \> 8 \>
$30$ \> ${A_{1,8}}^{5}{A_{3,16}}$ \> 8 \\
$30$ \> ${A_{1,8}}^{10}$ \> 10 \>
$32$ \> ${A_{2,9}}{C_{2,9}}{G_{2,12}}$ \> 6 \\
$32$ \> ${A_{2,9}}{A_{4,15}}$ \> 6 \>
$32$ \> ${A_{2,9}}^{4}$ \> 8 \\
$32$ \> ${A_{1,6}}{A_{3,12}}{G_{2,12}}$ \> 6 \>
$32$ \> ${A_{1,6}}{A_{2,9}}{C_{3,12}}$ \> 6 \\
$32$ \> ${A_{1,6}}{A_{2,9}}{B_{3,15}}$ \> 6 \>
$32$ \> ${A_{1,6}}^{2}{A_{2,9}}^{2}{C_{2,9}}$ \> 8 \\
$32$ \> ${A_{1,6}}^{3}{A_{2,9}}{A_{3,12}}$ \> 8 \>
$32$ \> ${A_{1,6}}^{4}{C_{2,9}}^{2}$ \> 8 \\
$32$ \> ${A_{1,6}}^{6}{G_{2,12}}$ \> 8 \>
$32$ \> ${A_{1,6}}^{8}{A_{2,9}}$ \> 10 \\
$36$ \> ${C_{4,10}}$ \> 4 \>
$36$ \> ${B_{4,14}}$ \> 4 \\
$36$ \> ${A_{3,8}}{C_{3,8}}$ \> 6 \>
$36$ \> ${A_{3,8}}{B_{3,10}}$ \> 6 \\
$36$ \> ${A_{2,6}}{G_{2,8}}^{2}$ \> 6 \>
$36$ \> ${A_{2,6}}{D_{4,12}}$ \> 6 \\
$36$ \> ${A_{2,6}}^{2}{C_{2,6}}^{2}$ \> 8 \>
$36$ \> ${A_{1,4}}{A_{2,6}}{A_{3,8}}{C_{2,6}}$ \> 8 \\
$36$ \> ${A_{1,4}}^{2}{C_{2,6}}^{3}$ \> 8 \>
$36$ \> ${A_{1,4}}^{2}{A_{3,8}}^{2}$ \> 8 \\
$36$ \> ${A_{1,4}}^{2}{A_{2,6}}^{2}{G_{2,8}}$ \> 8 \>
$36$ \> ${A_{1,4}}^{4}{C_{2,6}}{G_{2,8}}$ \> 8 \\
$36$ \> ${A_{1,4}}^{4}{A_{4,10}}$ \> 8 \>
$36$ \> ${A_{1,4}}^{4}{A_{2,6}}^{3}$ \> 10 \\
$36$ \> ${A_{1,4}}^{5}{C_{3,8}}$ \> 8 \>
$36$ \> ${A_{1,4}}^{5}{B_{3,10}}$ \> 8 \\
$36$ \> ${A_{1,4}}^{6}{A_{2,6}}{C_{2,6}}$ \> 10 \>
$36$ \> ${A_{1,4}}^{7}{A_{3,8}}$ \> 10 \\
$36$ \> ${A_{1,4}}^{12}$ \> 12 * \>
$40$ \> ${A_{1,3}}^{4}{G_{2,6}}^{2}$ \> 8 \\
$40$ \> ${A_{1,3}}^{4}{D_{4,9}}$ \> 8 \>
$42$ \> ${A_{2,4}}^{4}{C_{2,4}}$ \> 10 $\oplus$ \\
$48$ \> ${C_{2,3}}^{2}{G_{2,4}}^{2}$ \> 8 $\oplus$ \>
$48$ \> ${C_{2,3}}^{2}{D_{4,6}}$ \> 8 \\
$48$ \> ${A_{6,7}}$ \> 6 \>
$48$ \> ${A_{4,5}}{C_{2,3}}{G_{2,4}}$ \> 8 \\
$48$ \> ${A_{4,5}}^{2}$ \> 8 \>
$48$ \> ${A_{2,3}}{C_{2,3}}^{4}$ \> 10 \\
$48$ \> ${A_{2,3}}{A_{3,4}}^{2}{C_{2,3}}$ \> 10 \>
$48$ \> ${A_{2,3}}^{3}{C_{2,3}}{G_{2,4}}$ \> 10 \\
$48$ \> ${A_{2,3}}^{3}{A_{4,5}}$ \> 10 \>
$48$ \> ${A_{2,3}}^{6}$ \> 12 \\
$48$ \> ${A_{1,2}}{D_{5,8}}$ \> 6 \>
$48$ \> ${A_{1,2}}{C_{2,3}}{C_{3,4}}{G_{2,4}}$ \> 8 \\
$48$ \> ${A_{1,2}}{B_{3,5}}{C_{2,3}}{G_{2,4}}$ \> 8 \>
$48$ \> ${A_{1,2}}{A_{5,6}}{C_{2,3}}$ \> 8 \\
$48$ \> ${A_{1,2}}{A_{4,5}}{C_{3,4}}$ \> 8 \>
$48$ \> ${A_{1,2}}{A_{4,5}}{B_{3,5}}$ \> 8 \\
$48$ \> ${A_{1,2}}{A_{3,4}}{C_{2,3}}^{3}$ \> 10 \>
$48$ \> ${A_{1,2}}{A_{3,4}}^{3}$ \> 10 \\
$48$ \> ${A_{1,2}}{A_{2,3}}^{2}{A_{3,4}}{G_{2,4}}$ \> 10 \>
$48$ \> ${A_{1,2}}{A_{2,3}}^{3}{C_{3,4}}$ \> 10 \\
$48$ \> ${A_{1,2}}{A_{2,3}}^{3}{B_{3,5}}$ \> 10 \>
$48$ \> ${A_{1,2}}^{2}{G_{2,4}}^{3}$ \> 8 \\
$48$ \> ${A_{1,2}}^{2}{D_{4,6}}{G_{2,4}}$ \> 8 \>
$48$ \> ${A_{1,2}}^{2}{C_{3,4}}^{2}$ \> 8 \\
$48$ \> ${A_{1,2}}^{2}{B_{3,5}}{C_{3,4}}$ \> 8 \>
$48$ \> ${A_{1,2}}^{2}{B_{3,5}}^{2}$ \> 8 \\
$48$ \> ${A_{1,2}}^{2}{A_{2,3}}{C_{2,3}}^{2}{G_{2,4}}$ \> 10 \>
$48$ \> ${A_{1,2}}^{2}{A_{2,3}}{A_{4,5}}{C_{2,3}}$ \> 10 \\
$48$ \> ${A_{1,2}}^{2}{A_{2,3}}^{4}{C_{2,3}}$ \> 12 \>
$48$ \> ${A_{1,2}}^{3}{A_{3,4}}{C_{2,3}}{G_{2,4}}$ \> 10 \\
$48$ \> ${A_{1,2}}^{3}{A_{3,4}}{A_{4,5}}$ \> 10 \>
$48$ \> ${A_{1,2}}^{3}{A_{2,3}}{C_{2,3}}{C_{3,4}}$ \> 10 \\
$48$ \> ${A_{1,2}}^{3}{A_{2,3}}{B_{3,5}}{C_{2,3}}$ \> 10 \>
$48$ \> ${A_{1,2}}^{3}{A_{2,3}}^{3}{A_{3,4}}$ \> 12 \\
$48$ \> ${A_{1,2}}^{4}{C_{4,5}}$ \> 8 \>
$48$ \> ${A_{1,2}}^{4}{B_{4,7}}$ \> 8 \\
$48$ \> ${A_{1,2}}^{4}{A_{3,4}}{C_{3,4}}$ \> 10 \>
$48$ \> ${A_{1,2}}^{4}{A_{3,4}}{B_{3,5}}$ \> 10 \\
$48$ \> ${A_{1,2}}^{4}{A_{2,3}}{G_{2,4}}^{2}$ \> 10 \>
$48$ \> ${A_{1,2}}^{4}{A_{2,3}}{D_{4,6}}$ \> 10 \\
$48$ \> ${A_{1,2}}^{4}{A_{2,3}}^{2}{C_{2,3}}^{2}$ \> 12 \>
$48$ \> ${A_{1,2}}^{5}{A_{2,3}}{A_{3,4}}{C_{2,3}}$ \> 12 \\
$48$ \> ${A_{1,2}}^{6}{C_{2,3}}^{3}$ \> 12 \>
$48$ \> ${A_{1,2}}^{6}{A_{3,4}}^{2}$ \> 12 \\
$48$ \> ${A_{1,2}}^{6}{A_{2,3}}^{2}{G_{2,4}}$ \> 12 \>
$48$ \> ${A_{1,2}}^{8}{C_{2,3}}{G_{2,4}}$ \> 12 \\
$48$ \> ${A_{1,2}}^{8}{A_{4,5}}$ \> 12 \>
$48$ \> ${A_{1,2}}^{8}{A_{2,3}}^{3}$ \> 14 \\
$48$ \> ${A_{1,2}}^{9}{C_{3,4}}$ \> 12 \>
$48$ \> ${A_{1,2}}^{9}{B_{3,5}}$ \> 12 \\
$48$ \> ${A_{1,2}}^{10}{A_{2,3}}{C_{2,3}}$ \> 14 \>
$48$ \> ${A_{1,2}}^{11}{A_{3,4}}$ \> 14 \\
$48$ \> ${A_{1,2}}^{16}$ \> 16 * \>
$56$ \> ${G_{2,3}}^{4}$ \> 8 \\
$56$ \> ${C_{3,3}}^{2}{G_{2,3}}$ \> 8 \>
$60$ \> ${C_{2,2}}^{6}$ \> 12 * \\
$60$ \> ${A_{2,2}}{F_{4,6}}$ \> 6 $\dagger$ \>
$60$ \> ${A_{2,2}}^{4}{D_{4,4}}$ \> 12 \\
$60$ \> ${A_{2,2}}^{5}{C_{2,2}}^{2}$ \> 14 \>
$72$ \> ${A_{3,2}}^{2}{G_{2,2}}^{3}$ \> 12 \\
$72$ \> ${A_{3,2}}^{2}{D_{4,3}}{G_{2,2}}$ \> 12 \>
$72$ \> ${A_{3,2}}^{2}{C_{3,2}}^{2}$ \> 12 \\
$72$ \> ${A_{1,1}}{C_{5,3}}{G_{2,2}}$ \> 8 \>
$72$ \> ${A_{1,1}}^{2}{D_{6,5}}$ \> 8 \\
$72$ \> ${A_{1,1}}^{2}{C_{3,2}}{D_{5,4}}$ \> 10 \>
$72$ \> ${A_{1,1}}^{2}{A_{3,2}}^{3}{C_{3,2}}$ \> 14 \\
$72$ \> ${A_{1,1}}^{3}{C_{3,2}}{G_{2,2}}^{3}$ \> 12 \>
$72$ \> ${A_{1,1}}^{3}{C_{3,2}}{D_{4,3}}{G_{2,2}}$ \> 12 \\
$72$ \> ${A_{1,1}}^{3}{C_{3,2}}^{3}$ \> 12 \>
$72$ \> ${A_{1,1}}^{3}{A_{7,4}}$ \> 10 \\
$72$ \> ${A_{1,1}}^{3}{A_{5,3}}{G_{2,2}}^{2}$ \> 12 \>
$72$ \> ${A_{1,1}}^{3}{A_{5,3}}{D_{4,3}}$ \> 12 \\
$72$ \> ${A_{1,1}}^{4}{A_{3,2}}{D_{5,4}}$ \> 12 \>
$72$ \> ${A_{1,1}}^{4}{A_{3,2}}^{4}$ \> 16 * \\
$72$ \> ${A_{1,1}}^{5}{A_{3,2}}{G_{2,2}}^{3}$ \> 14 \>
$72$ \> ${A_{1,1}}^{5}{A_{3,2}}{D_{4,3}}{G_{2,2}}$ \> 14 \\
$72$ \> ${A_{1,1}}^{5}{A_{3,2}}{C_{3,2}}^{2}$ \> 14 \>
$72$ \> ${A_{1,1}}^{7}{A_{3,2}}^{2}{C_{3,2}}$ \> 16 \\
$72$ \> ${A_{1,1}}^{9}{D_{5,4}}$ \> 14 \>
$72$ \> ${A_{1,1}}^{9}{A_{3,2}}^{3}$ \> 18 \\
$72$ \> ${A_{1,1}}^{10}{G_{2,2}}^{3}$ \> 16 \>
$72$ \> ${A_{1,1}}^{10}{D_{4,3}}{G_{2,2}}$ \> 16 \\
$72$ \> ${A_{1,1}}^{10}{C_{3,2}}^{2}$ \> 16 \>
$72$ \> ${A_{1,1}}^{12}{A_{3,2}}{C_{3,2}}$ \> 18 \\
$72$ \> ${A_{1,1}}^{14}{A_{3,2}}^{2}$ \> 20 \>
$72$ \> ${A_{1,1}}^{17}{C_{3,2}}$ \> 20 \\
$72$ \> ${A_{1,1}}^{19}{A_{3,2}}$ \> 22 \>
$72$ \> ${A_{1,1}}^{24}$ \> 24 * \\
$84$ \> ${B_{3,2}}^{4}$ \> 12 * \>
$84$ \> ${A_{4,2}}^{2}{C_{4,2}}$ \> 12 \\
$96$ \> ${C_{2,1}}^{4}{D_{4,2}}^{2}$ \> 16 * \>
$96$ \> ${A_{2,1}}{C_{2,1}}{E_{6,4}}$ \> 10 \\
$96$ \> ${A_{2,1}}{C_{2,1}}^{6}{D_{4,2}}$ \> 18 \>
$96$ \> ${A_{2,1}}^{2}{D_{4,2}}{F_{4,3}}$ \> 12 \\
$96$ \> ${A_{2,1}}^{2}{C_{2,1}}^{8}$ \> 20 \>
$96$ \> ${A_{2,1}}^{2}{A_{8,3}}$ \> 12 \\
$96$ \> ${A_{2,1}}^{2}{A_{5,2}}^{2}{C_{2,1}}$ \> 16 \>
$96$ \> ${A_{2,1}}^{3}{C_{2,1}}^{2}{F_{4,3}}$ \> 14 \\
$96$ \> ${A_{2,1}}^{5}{D_{4,2}}^{2}$ \> 18 \>
$96$ \> ${A_{2,1}}^{6}{C_{2,1}}^{2}{D_{4,2}}$ \> 20 \\
$96$ \> ${A_{2,1}}^{7}{C_{2,1}}^{4}$ \> 22 \>
$96$ \> ${A_{2,1}}^{12}$ \> 24 * \\
$108$ \> ${B_{4,2}}^{3}$ \> 12 * \>
$120$ \> ${E_{6,3}}{G_{2,1}}^{3}$ \> 12 \\
$120$ \> ${C_{3,1}}^{2}{E_{6,3}}$ \> 12 \>
$120$ \> ${A_{3,1}}{D_{7,3}}{G_{2,1}}$ \> 12 \\
$120$ \> ${A_{3,1}}{C_{7,2}}$ \> 10 \>
$120$ \> ${A_{3,1}}{C_{3,1}}{G_{2,1}}^{6}$ \> 18 \\
$120$ \> ${A_{3,1}}{C_{3,1}}^{3}{G_{2,1}}^{3}$ \> 18 \>
$120$ \> ${A_{3,1}}{C_{3,1}}^{5}$ \> 18 \\
$120$ \> ${A_{3,1}}{A_{7,2}}{G_{2,1}}^{3}$ \> 16 \>
$120$ \> ${A_{3,1}}{A_{7,2}}{C_{3,1}}^{2}$ \> 16 \\
$120$ \> ${A_{3,1}}^{2}{D_{5,2}}^{2}$ \> 16 * \>
$120$ \> ${A_{3,1}}^{5}{D_{5,2}}$ \> 20 \\
$120$ \> ${A_{3,1}}^{8}$ \> 24 * \>
$132$ \> ${A_{8,2}}{F_{4,2}}$ \> 12 \\
$144$ \> ${C_{4,1}}^{4}$ \> 16 * \>
$144$ \> ${B_{3,1}}^{2}{C_{4,1}}{D_{6,2}}$ \> 16 * \\
$144$ \> ${A_{4,1}}{B_{3,1}}^{4}{C_{4,1}}$ \> 20 \>
$144$ \> ${A_{4,1}}{A_{9,2}}{B_{3,1}}$ \> 16 \\
$144$ \> ${A_{4,1}}^{3}{C_{4,1}}^{2}$ \> 20 \>
$144$ \> ${A_{4,1}}^{6}$ \> 24 * \\
$156$ \> ${B_{6,2}}^{2}$ \> 12 * \>
$168$ \> ${D_{4,1}}^{6}$ \> 24 * \\
$168$ \> ${A_{5,1}}{E_{7,3}}$ \> 12 \>
$168$ \> ${A_{5,1}}{C_{5,1}}{E_{6,2}}$ \> 16 $\oplus$ \\
$168$ \> ${A_{5,1}}^{4}{D_{4,1}}$ \> 24 * \>
$192$ \> ${B_{4,1}}{C_{6,1}}^{2}$ \> 16 \\
$192$ \> ${B_{4,1}}^{2}{D_{8,2}}$ \> 16 * \>
$192$ \> ${A_{6,1}}{B_{4,1}}^{4}$ \> 22 \\
$192$ \> ${A_{6,1}}^{4}$ \> 24 * \>
$216$ \> ${A_{7,1}}{D_{9,2}}$ \> 16 * \\
$216$ \> ${A_{7,1}}^{2}{D_{5,1}}^{2}$ \> 24 * \>
$240$ \> ${C_{8,1}}{F_{4,1}}^{2}$ \> 16 \\
$240$ \> ${B_{5,1}}{E_{7,2}}{F_{4,1}}$ \> 16 \>
$240$ \> ${A_{8,1}}^{3}$ \> 24 * \\
$264$ \> ${D_{6,1}}^{4}$ \> 24 * \>
$264$ \> ${A_{9,1}}^{2}{D_{6,1}}$ \> 24 * \\
$288$ \> ${B_{6,1}}{C_{10,1}}$ \> 16 \>
$300$ \> ${B_{12,2}}$ \> 12 * \\
$312$ \> ${E_{6,1}}^{4}$ \> 24 * \>
$312$ \> ${A_{11,1}}{D_{7,1}}{E_{6,1}}$ \> 24 * \\
$336$ \> ${A_{12,1}}^{2}$ \> 24 * \>
$360$ \> ${D_{8,1}}^{3}$ \> 24 * \\
$384$ \> ${B_{8,1}}{E_{8,2}}$ \> 16 $\dagger$ \>
$408$ \> ${A_{15,1}}{D_{9,1}}$ \> 24 * \\
$456$ \> ${D_{10,1}}{E_{7,1}}^{2}$ \> 24 * \>
$456$ \> ${A_{17,1}}{E_{7,1}}$ \> 24 * \\
$552$ \> ${D_{12,1}}^{2}$ \> 24 * \>
$624$ \> ${A_{24,1}}$ \> 24 * \\
$744$ \> ${E_{8,1}}^{3}$ \> 24 * \>
$744$ \> ${D_{16,1}}{E_{8,1}}$ \> 24 * \\
$1128$ \> ${D_{24,1}}$ \> 24
\end{tabbing}
\section{Uniqueness of the $N=0$ theory}
\label{uniqueness}
It is to be expected, by analogy with the situation for codes and also
lattices, that the $N=0$ theory would be unique. (Note that such analogies
are not perfect, since in Venkov's work on the even self-dual lattices,
which the previous section mirrored, there is one and only one lattice
corresponding to each possible combination of algebras, whereas Schellekens
has observed that there are some algebras in the above list for which it
is impossible to obtain even a modular invariant combination of characters,
let alone a fully consistent conformal field theory. Nevertheless, we expect
uniqueness when the theories do exist, and almost certainly uniqueness is
to be expected in the case $N=0$ [in the case of lattices, the uniqueness
proofs for $N=0$ and $N>0$ are distinct]). Let us consider this problem.

Suppose that $\Hil$ is a self-dual $c=24$ conformal field theory with
no weight one states. Suppose an involution $g$ exists,
and consider the
orbifold constructed using it (which we also suppose to exist).
We may evaluate the partition function for the orbifold in terms of the
Thompson series for the involution, {\it i.e.}
\begin{eqnarray}
\chi_{\Hil_g}(\tau)&=&{1\over 2}\left(
{1\lower 6pt \hbox{$
{\square\atop{\displaystyle 1}}$}}
+{g\lower 6pt \hbox{$
{\square\atop{\displaystyle 1}}$}}
+{1\lower 6pt \hbox{$
{\square\atop{\displaystyle g}}$}}
+{g\lower 6pt \hbox{$
{\square\atop{\displaystyle g}}$}}\right) \\
&=&{1\over 2}\left(J(\tau)+T_g(\tau)+T_g(S(\tau))+T_g(ST(\tau))
\right)\,.
\end{eqnarray}
Clearly, $T_g(\tau)={g\lower 6pt \hbox{$
{\square\atop{\displaystyle 1}}$}}$ is invariant under $\Gamma_0(2)$,
{\it i.e.} under $T$ and $ST^2S$. If it has the correct behaviour at
$q=1$  ({\it i.e.} if the ground state of the twisted sector has
energy $\geq 1$ \cite{Tuite:moon})
then it is a $\Gamma_0(2)$ hauptmodul.
Thus, it is known explicitly:
\begin{equation}
T_g(\tau)=\left({\eta(\tau)\over{\eta(2\tau)}}\right)^{24}+24\,.
\end{equation}
We find that
\begin{equation}
\chi_{\Hil_g}(\tau)=J(\tau)+24\,.
\end{equation}
Thus, in the above notation $N=24$, and we see from Schellekens' list
that the only possibility is to have algebra $U(1)^{24}$. Now,
we have the theorem from \cite{DGMtriality} that
when the rank is equal to the central
charge, the theory is equivalent to an FKS lattice theory $\Hil(\Lambda)$ for
some even lattice $\Lambda$, and further that the CFT is self-dual if and only
if the lattice is self-dual.
So $\Hil_g\cong\Hil(\Lambda_{24})$,
the Leech lattice being the unique even self-dual lattice in 24 dimensions
having no vectors of length squared two (thus giving 24 weight 1 states in the
CFT).

To proceed now in a way analogous to Venkov's proof of the uniqueness of
the Leech lattice ({\em i.e.} following the spirit of Schellekens' approach
in the previous section), we would assume the existence of an automorphism
$h$ of our new theory $\Hil_g$ such that $(\Hil_g)_h$ exists and is
isomorphic to the original theory $\Hil$, {\em i.e.} we assume the
existence of an inverse to the orbifold construction.
The projection onto $h$ invariant states removes all the
weight one states, by definition. Hence we must have that $h$ acts as
$-1$ on the states $a_{-1}^j|0\rangle$ (using the notation of
\cite{DGMtwisted}).
We deduce from this
that $h$ is simply the automorphism of $\Hil(\Lambda_{24})$ induced by
the reflection twist on $\Lambda_{24}$. But we know that $\Hil(\Lambda_
{24})_{-1}\cong V^\natural$, the natural Monster
module\cite{FLMbook,DGMtriality}.
Hence, $\Hil\cong V^\natural$
as required.

Note that this is {\em not} a proof of the uniqueness of the $N=0$ theory. What
we
have demonstrated is that the uniqueness problem, modulo some general
results on orbifold theory not specific to the Monster, is equivalent
to the hauptmodul property, or, as Tuite has shown, to the nature of the
energy and degeneracy
of the twisted sector ground state. This exercise has merely demonstrated
how Schellekens' work allows us to tackle problems by
giving us data on the possible $c=24$ theories. What we have done
is essentially analogous to Venkov's proof of the uniqueness of the
Monster in the same way that Schellekens' work itself is simply an analogue
of (part of) Venkov's reformulation of
Niemeier's classification of lattices.
The power of Schellekens' result is that, we can say what the theory
$\Hil_g$ must be without an explicit construction (though such a construction
would
of course be needed to demonstrate the actual consistency of
the orbifold theory).
The problem, as always, is that we lack any general formulation of
orbifold construction, but can only do it in ``simple" cases such as in
\cite{DGMtwisted}.

If, on the other hand, we do not want to follow Venkov's method
too closely, then we can provide a more convincing ``proof"
of the uniqueness of the $N=0$ conformal field theory.
We have that $\Hil_g=\Hil^0_g\oplus U$,
with $\Hil^0_g$ the
invariant sector of $\Hil$ under the action of $g$ and $U$ a
representation of $\Hil_g^0$.
Now, the states $a_{-1}^j|0\rangle$ lie
in $U$, and so the states $a_{-m}^ia_{-n}^j|0\rangle$ lie in
$\Hil_g^0$. Thus, we see that $\Hil^0_g\cong\Hil(\Lambda_{24})_+$,
that part of $\Hil(\Lambda_{24})$ invariant under the reflection twist
introduced in\cite{DGMtwisted}
(which we can easily check is consistent, as we know its partition
function is ${1\over 2}\left({1\lower 6pt \hbox{$
{\square\atop{\displaystyle 1}}$}}+
{g\lower 6pt \hbox{$
{\square\atop{\displaystyle 1}}$}}\right)={1\over 2}\left(
J(\tau)+T_g(\tau)\right)$.).
We have proved\cite{DGMtriality}
that there are only two representations of $U$ of $\Hil(\Lambda)_+$ for
$\Lambda$
self-dual such that $\Hil(\Lambda)_+\oplus U$ is
a consistent (self-dual) conformal field theory. These representations are
distinguished by the number of weight one states. So we must have
that $\Hil\cong V^\natural$, as required. Note that this proof circumvents
the need to invoke
any argument about inverting an orbifold construction, though still
relies upon the assumption of the existence of a suitable involution $g$ of
the original theory.

The same technique may be used instead beginning with a higher order
automorphism, though of course we would then need to know about higher order
twisted constructions of the Monster conformal field theory. This is
work which is still in progress\cite{PSMthird}]
\section{Complementary representations}
Consider a bosonic meromorphic hermitian conformal field theory $S$ which
may be extended to form a self-dual theory by adding in a representation $U$
(real, hermitian and satisfying the additional locality requirement as detailed
in \cite{thesis,DGMtriality}). Thus
${\cal H}=S\oplus U$ is self-dual.
\subsection{Definition}
Now, let $\theta$ be the automorphism of ${\cal H}$ defined to be $1$ on $S$
and
$-1$ on $U$. The invariant sub-theory is simply $S$, and we shall assume that
we may construct a corresponding self-dual orbifold theory
${\cal H}_\theta=S\oplus
U'$, with $U'$ a representation of $S$.

We have
\begin{equation}
\chi_{{\cal H}_\theta}(\tau)=\chi_S(\tau)+\chi_{U'}(\tau)\,,
\end{equation}
where
\begin{eqnarray}
\chi_{U'}(\tau)&=&{1\over 2}\left(
{1\lower 6pt \hbox{$
{\square\atop\displaystyle\theta}$}}(\tau)+
{\theta\lower 6pt \hbox{$
{\square\atop\displaystyle\theta}$}}(\tau)\right) \nonumber \\
\chi_S(\tau)&=&{1\over 2}\left(\chi_\Hil(\tau)+{\theta
\lower 6pt \hbox{$
{\square\atop{\displaystyle 1}}$}}(\tau)\right)\\
\end{eqnarray}
and all the boxes are to be understood with reference to $\Hil$.

Thus
\begin{equation}
\chi_S(S(\tau))={1\over 2}\left(\chi_\Hil(\tau)+{1
\lower 6pt \hbox{$
{\square\atop\displaystyle\theta}$}}(\tau)\right)
\end{equation}
and
\begin{equation}
\chi_S(ST(\tau))={1\over 2}\left(\chi_\Hil(\tau)+{\theta
\lower 6pt \hbox{$
{\square\atop\displaystyle\theta}$}}(\tau)\right)\,.
\end{equation}
Hence
\begin{equation}
\chi_\Hil(\tau)+\chi_{{\cal H}_\theta}(\tau)=\chi_S(\tau)+\chi_S(S(\tau))+
\chi_S(ST(\tau))\,.
\label{sum}
\end{equation}
In other words, the sum of the partition functions of $\Hil$ and
$\Hil_\theta$ is determined solely
in terms of the partition function of the invariant sub-theory $S$.
For $c=24$, the partition functions are restricted to be of the form
$J(\tau)+N$,
where $J$ is the elliptic modular function with zero constant term and $N$ is
the
number of weight one states. So
the above result simply states that the sum of the number
of weight one states in $\Hil$ and $\Hil_\theta$ is determined solely by the
partition function of $S$.

\proclaim Definition. We shall say that $U'$ is the complementary
representation
of $S$ to $U$. Further, if $U$ and $U'$ are equivalent as representations of
$S$,
we say that $U$ is self-complementary with respect to $S$. \par

\subsection{Example}
Consider $S=\Hil(\Lambda)_+$, $\Lambda$ an even self-dual
lattice. We have shown in \cite{DGMtriality} that there exist two
representations
$U_1$ and $U_2$ extending $S$ to a self-dual $c=24$ conformal field theory,
{\em i.e.} $\Hil(\Lambda)\cong S\oplus U_1$ and $\widetilde\Hil(\Lambda)
)\cong S\oplus U_2$.
Consider taking $U=U_1$ in the above notation, and construct $U'$.
We can argue that ${U_1}'\cong U_2$. Suppose not. Then $U_1$ must be
self-complementary, and so the number of weight one states in $U_1$
would then be fixed by the partition function of $S$. Now, assuming that the
notion of complementarity is symmetric,{\em i.e.} $U''\cong U$
(see below for more discussion),
$U_2$ must also be self-complementary (otherwise ${U_2}'\cong U_1$, and hence
${U_1}'\cong{U_2}''\cong U_2$). So \reg{sum} then implies that $U_2$ must
have the same number of weight one states as $U_1$. We know from the
explicit constructions of these representations that this is a contradiction,
as required.

Note the advantage of this point of view is that it puts $\Hil(\Lambda)$
and $\widetilde\Hil(\Lambda)$ on an equal footing, unlike the conventional
approach.
\subsection{Symmetry of the definition}
We discuss here the property alluded to in the above example.
Suppose $U'$ is the complementary representation of the conformal field theory
$S$ to a
representation $U$. We can then construct a representation $U''$ of S
complementary to $U'$. We see from \reg{sum} that it must have the
same partition function as $U$ (which is more than saying merely
that the number of weight one states is the same, at least if
$c>24$). This observation alone is enough in the case of our example above,
since we know that $U_1$ and $U_2$ have distinct partition functions and
so symmetry is assured. The question in general of whether $U\cong U''$
remains an open one however. Even for $c=24$, we have examples of
distinct theories with the same number of weight one states, and so
the equality of the partition functions alone is not sufficient, although
it is a conjecture which we still expect to be true.
\subsection{Application to self-dual $c=24$ conformal field theories}
\label{applic}
We may ask exactly what we must know of the partition function of the theory
$S$
in order to calculate the sum in \reg{sum}. Restricting our considerations
to central charge 24, we need only worry about the number of states
of conformal weight one.

Clearly, $\chi_S(\tau)$ is a $\Gamma_0(2)$ invariant. So we can
write\cite{Tuite:moon}
\begin{equation}
\chi_S(\tau)=N_S+{1\over 2}\left\{J(\tau)+\alpha\left(
{\theta_3(\tau)^8\theta_4(\tau)^8+2^{-4}\theta_2(\tau)^{16}\over
{\eta(\tau)^8\eta(2\tau)^8}}+24\right)
+\beta\left(\left({\eta(\tau)\over{\eta(2\tau)}}\right)^{24}+24\right)\right\}\,,
\end{equation}
using the $\Gamma_0(2)$ and $\Gamma_0(2)+$ hauptmoduls,
where $N_S$ is the number of states of weight 1 in $S$ and $\alpha+\beta=1$
(since $\chi_S(\tau)\sim q^{-1}$ as $q\rightarrow 0$).
(Note that if $\alpha=0$ or $1$ then
the Moonshine
Conjecture\cite{moonshine} holds in this case.)
Considering the transformation $S:\tau\mapsto-1/\tau$ shows that
$\alpha$ is to be interpreted as the number of weight ${1\over 2}$ states in
the
twisted sector, and is given by
\begin{equation}
\alpha={N_{S,2}-98580\over{2048}}\,,
\label{alpha}
\end{equation}
where $N_{S,2}$ is the number of states of weight 2 in $S$.
We also find, using \reg{sum}, that
\begin{equation}
N_\Hil+N_{\Hil'}=3N_S+24(1-\alpha)\,,
\label{beta}
\end{equation}
where $N_\Hil$ is the number of weight one states in $\Hil=S\oplus
U$, and similarly for $N_{\Hil'}$.

In the following sections, we will apply our arguments to cases in which
the theory $S$ is constructed as
the invariant sub-theory of a
self-dual theory under an involution, and we are thus simply
considering the
orbifold of the first theory with respect to this involution.
The representations of the sub-theory giving us the orbifold and
the original theory are complementary, allowing us to use
\reg{alpha} and \reg{beta} to calculate the number of weight one states in the
orbifold theory.
Together with a knowledge of the Kac-Moody algebra corresponding to
the weight one states of $S$, the number
of weight one states allows us to look up in the table given by the
results of Schellekens in section \ref{Schell} and identify the most
likely candidate for the theory.
An absence of a suitable candidate will imply that the orbifold is not
consistent.

[The motivation for the above approach is based upon an extension of
the analogies between constructions of lattices from binary codes and
constructions of CFT's from lattices which were summarised in
\cite{DGMtrialsumm}. In \cite{PSMcodes} constructions for all of the
Niemeier lattices from ternary codes were given, suggesting the
existence of corresponding constructions of CFT's from (Eisenstein)
lattices by some form of $\ze_2$-orbifold approach. The exact nature
of the orbifold theory would be difficult to write down. Instead it is
considerably easier to postulate some sub-theory with $c=24$ of the
$c=48$ theory
$\Hil(\Lambda)$ ($\Lambda$ the 48-dimensional even self-dual lattice
corresponding to a 24-dimensional Eisenstein lattice) which is to form
the invariant space upon which the
orbifold is constructed. The above argument would then give us the sum
of the number of weight one states in two (potentially distinct)
orbifolds that may be then formed by the addition of complementary
representations, and reference to the Kac-Moody
algebra of the invariant theory and Schellekens' results would provide
information hopefully sufficient to identify the theories. This program
is still in progress.

In any case, note that \reg{sum} is analogous to the situation we have in
constructing lattices from ternary codes in \cite{PSMcodes}.
In that case we had
a lattice $\Lambda_\C^0$ and formed a pair $\Lambda_\C^\pm$ by adding
appropriate vectors. The final number of length squared 2 vectors
in the even self-dual lattices $\Lambda_\C^0\cup\Lambda_\C^\pm$ turned
out to be $3n_3+n_6+n_{24}^\pm$ respectively, the $3n_3+n_6$ coming from
$\Lambda_\C^0$ ($n_m$ is the number of codewords of weight $m$-see
\cite{PSMcodes} for a full description of the notation).
In general, $n_{24}^\pm$ are independent of $n_3$ and $n_6$,
but $n_{24}$ is not\cite{PSMcodes}, {\em i.e.}
$|\Lambda_\C^+(2)|+|\Lambda_\C^-(2)|$ is fixed by $\Lambda_\C^0$.
Similarly, we note that \reg{beta} is again analogous to the lattice
situation, for which we
have
$|\Lambda_\C^+(2)|+|\Lambda_\C^-(2)|=3|\Lambda_\C^0(2)|+48(1-
{1\over 2}n_3)$.]
\subsection{Application to the theories $\Hil(\Lambda)$}
Let us apply the above considerations to some simple examples. Consider
projections by arbitrary involutions of the Niemeier lattice theories
$\Hil(\Lambda)$.
Schellekens and Yankielowicz in \cite{SchellYank:curious}
have already observed that this is a useful thing to consider, since
one of their two proposed new theories they claimed could be regarded
as a $\ze_2$-orbifold of the theory $\Hil({E_8}^3)$ induced by the
involution on the lattice which interchanges two of the $E_8$ factors and
shifts the third by a $D_8$ weight vector (although we feel that they
should
really consider a reflection on the third factor instead,
since the shift is {\em not} a lattice automorphism!).

Note that it seems to be known, at least to the
mathematicians\cite{Mythesis}, how to construct the twisted vertex
operators corresponding to an arbitrary lattice automorphism. For an
involution, the techniques applied in \cite{DGMtwisted} may then be
used to construct the corresponding intertwining vertex operators
which enable one to unite the invariant theory and its representation
into a consistent CFT. The question of the consistency of such a
theory still has to be resolved by direct calculation though,
analogous to that carried out in \cite{DGMtwisted}. In any case, we
can certainly calculate the ground state energy and degeneracy of the
twisted sector, assuming consistency. Thus, use of the above
techniques may seem unnecessary, though they do provide a quicker
route to the answer and are really the only tool which can be used in
the more complex situation considered in the next subsection where no
explicit construction of the orbifold is yet known.

We need to consider the automorphism groups of the Niemeier
lattices\cite{ConSlo}.
The lattices are specified by giving
the root system and specifying a set of glue vectors, {\em i.e.} a set
of vectors in the dual of the root system which, taken together with
the root system, span the Niemeier lattice. The automorphism
groups are composed of three pieces. One permutes the different
components in the root system. One permutes the glue within a given
component but leaves the components fixed, while the third leaves both
the glue and the components fixed. Note that we shall frequently refer
to the Niemeier lattices simply by specifying the semi-simple Lie
algebras corresponding to their root system. This should not be
confused with the root lattice of the algebra.

Note that not all involutions give rise to a consistent orbifold
theory. All cases of this which we will come across can be
eliminated simply by observing that the energy of the twisted sector
ground state is not in ${1\over 2}\ze$. We will see an example of this
in constructing our chosen example.

Let us consider an example.
The glue code for the Niemeier lattice ${E_6}^4$ is the tetracode
$\C_4$\cite{ConSlo}.
This has an involution of the form $(\leftrightarrow 1 -1)$,
using the obvious notation.  If we try to take this as our involution,
it gives a twist invariant algebra ${E_6}^2C_4$, and the argument of
section \ref{applic} tells us that the number of weight one states in
the orbifold theory is $288-24\alpha$ (we can work out $\alpha$ if
required). The only possible theory from Schellekens' list of
sufficient rank would need $\alpha=5$ and have algebra $A_5C_5E_6$,
which is inconsistent. The reason for failure is that the
ground state energy of the twisted sector is $6/16$
>from the $-1$ on one $E_6$ plus $6/16$ from the interchange of the pair
of $E_6$'s, which is not half-integral.

Let us try replacing the $1$ in the specification of the automorphism
with an involution of $E_6$ (which leaves the glue fixed). We try a
one with invariant algebra $A_5A_1$\cite{Mythesis}, which we
can check leaves the glue unaffected. Thus, the twist
invariant algebra is $E_6A_5A_1C_4$, and we find the number of weight
one states to be $168-24\alpha$. This is consistent with the {\em new
theory} $A_5C_5E_{6,2}$, if we demonstrate that $\alpha=0$. Note that,
>from \cite{Mythesis}, we see that the extra automorphism in this case
contributes $1/4$ to
the ground state energy in the twisted sector, making it integral (and
also incidentally implying that $\alpha=0$, since there are no weight
$1/2$ states). Also, note that the weight one state argument implies
that 16 new weight one states must come from the twisted sector, as
an explicit construction along the lines discussed in \cite{Mythesis}
would verify.

One may proceed systematically through the Niemeier lattices and the
involutions of each. We present ${E_8}^3$ as the simplest example.
Note that we merely consider the lattice involutions, whereas the
automorphism group of the conformal field
theory is extended by the presence of the
cocycles. The implication of the presence of additional involutions
due to these must also be taken into account. We will discuss this
briefly below.
We have three possibilities for contributions to the lattice
involution of ${E_8}^3$. We have transposition of a pair of the
$E_8$'s and also two involutions of the $E_8$ root lattice itself. We
refer to these as $\theta_1$ and $\theta_2$. $\theta_1$ is simply the
reflection, and gives a contribution of $1/2$ to the twisted sector
ground state energy with degeneracy 16 (see for example \cite{DGMtwisted}).
The corresponding invariant algebra is $D_8$. $\theta_2$ has invariant
subalgebra $E_7A_1$, and contributes $1/4$ to the energy with
degeneracy 2\cite{Mythesis}.

The results are tabulated in table \ref{e8inv}. We label the orbifold
by the algebra, though of course we only know the corresponding theory
to (exist and) be unique in the rank 24 case\cite{DGMtriality}.
We see from applying our above arguments to the calculation of
$\alpha$ that we may interpret the transposition of two components of
the lattice as contributing $1/2$ to the energy of the ground state in
the twisted sector with unit degeneracy (independent of whether we
attach another automorphism). Note that theory 3 would give
$N=432-24\alpha$, with a subalgebra $E_8E_7A_1$. We can eliminate this
without even bothering to calculate $\alpha$ as we see that there is
no suitable theory in Schellekens' list. Note also that we can
eliminate it immediately since it would require a ground state energy
of $3/4$ in the twisted sector. This appears to be characteristic of
the theories, {\em i.e.} that the unusual or even inconsistent looking
algebras can be eliminated simply by consideration of the ground state
energy. Note that in some places we use known results for the
reflection twist\cite{DGMtrialsumm}, {\em i.e.} for the involutions 8
and 11, whereas for
1, 10, 13 and 16 we do not have sufficient knowledge from these simple
calculations to identify the conformal field theory completely.
Thus, we obtain the four distinct theories $\Hil(D_{10}{E_7}^2)$,
$\Hil({E_8}^3)$, $\Hil(E_8D_{16})$ and $\Hil({D_8}^3)$ and at least
one theory with algebra $E_{8,2}B_{8,1}$ by orbifolding $\Hil({E_8}^3)$
with respect
to the lattice induced involutions.

\begin{table}[htb]
\begin{center}
\begin{tabular}{|c|c|c|c|}\hline
Label & Involution & Orbifold & $\alpha$ \\ \hline
1 & $\leftrightarrow\cdot$ & ${E_8}^3$ or $E_8D_{16}$ & 1 \\
2 & $\leftrightarrow\theta_1$ & $E_{8,2}B_{8,1}$ & 0 \\
3 & $\leftrightarrow\theta_1$ & incorrect vacuum energy & $-$ \\
4 & $\theta_1\theta_2\theta_2$ & $D_{10,1}{E_{7,1}}^2$ & 0 \\
5 & $\theta_1\theta_1\theta_2$ & incorrect vacuum energy & $-$ \\
6 & $\theta_1\theta_1\theta_1$ & ${D_8}^3$ & 0 \\
7 & $\theta_2\theta_2\theta_2$ & incorrect vacuum energy & $-$ \\
8 & $\cdot\ \theta_1\theta_1$ & $E_8D_{16}$ & 0 \\
9 & $\cdot\ \theta_1\theta_2$ & incorrect vacuum energy & $-$ \\
10 & $\cdot\ \theta_2\theta_2$ & ${E_8}^3$ or $E_8D_{16}$ & 4 \\
11 & $\cdot\cdot\theta_1$ & ${E_8}^3$ & 16  \\
12 & $\cdot\cdot\theta_2$ & incorrect vacuum energy & $-$ \\
13 & $(\theta_1\leftrightarrow\theta_1)\ \cdot$ & ${E_8}^3$ or
$E_8D_{16}$ & 1 \\
14 & $(\theta_1\leftrightarrow\theta_1)\theta_1$ & $E_{8,2}B_{8,1}$ & 0 \\
15 & $(\theta_1\leftrightarrow\theta_1)\theta_2$ & incorrect vacuum
energy & $-$ \\
16 & $(\theta_2\leftrightarrow\theta_2)\ \cdot$ & ${E_8}^3$ or
$E_8D_{16}$ & 1 \\
17 & $(\theta_2\leftrightarrow\theta_2)\theta_1$ & $E_{8,2}B_{8,1}$ & 0 \\
18 & $(\theta_2\leftrightarrow\theta_2)\theta_2$ & incorrect vacuum
energy & $-$ \\ \hline
\end{tabular}
\caption{Involutions of ${E_8}^3$ and the corresponding orbifolds of
$\Hil({E_8}^3)$.}
\label{e8inv}
\end{center}
\end{table}

We may briefly consider the extension of the automorphism group of the
lattice due to the presence of the cocycles. We have automorphisms
$u a_n^j u^{-1}=R_{ij} a_n^i$,
$u|\lambda\rangle=(-1)^{\lambda\cdot\mu}|R\lambda\rangle$, $R\in{\rm
Aut}\ (\Lambda)$, $\mu\in\Lambda/2\Lambda$. Take the simple case
$R=1$. We use the construction for the root lattice $E_8$ given in the
appendix of Myhill's thesis. Then there are two inequivalent choices
for $\mu$, {\em i.e.} $e_1+e_2$ and ${1\over 2}\sum_{i=1}^8e_i$. These
are both found to give 136 invariant states, {\em i.e.} they behave
in the same way as $\theta_2$ on each component.

Let us consider what we may say about the twisted sector from our
point of view, even though we know its structure explicitly from
\cite{Mythesis}.
We find that, in
the cases where $\alpha=0$ (which appear to be most common), that the
partition function is of the form
$2^{12}\eta(\tau)/\eta(\tau/2)+M$, where $M$ is the number of
weight one states, {\em i.e.} it appears that we have $M$ states at
weight 1 with no descendants, and then the usual $2^{12}$ dimensional
ground state at level $3/2$ with the remainder of the states being
created by the action of 24 half-integrally graded bosonic creation
operators on these. This situation of course seems ridiculous. Let us
concentrate on the theory corresponding to involution 2 of table
\ref{e8inv}, {\em i.e.} the ``new'' Schellekens
theory\cite{SchellYank:curious}, and try to resolve this.

We have the partition function in the twisted sector as
\begin{equation}
{1\lower 6pt \hbox{$
{\square\atop{\displaystyle \theta}}$}}(\tau)=2^{12}\left(
{\eta(\tau)\over{\eta(\tau/2)}}\right)^{24}+16\,.
\label{FYA}
\end{equation}
However, we know that the $\theta_1$ part of the involution would be
expected to give a contribution of $2^4\left(
{\eta(\tau)\over{\eta(\tau/2)}}\right)^8q^{1/3}$ to this.
Considering the first few terms in the expansion of the ratio of the
two, we arrive at the conjecture
\begin{equation}
{1\lower 6pt \hbox{$
{\square\atop{\displaystyle \theta}}$}}(\tau)=2^4\left(
{\eta(\tau)\over{\eta(\tau/2)}}\right)^8\cdot
{\theta_{E_8}(\tau/2)\over{\eta(\tau/2)^8}}\,,
\label{twispart}
\end{equation}
which is trivial to check.

We postulate that the twisted sector is composed of a degeneracy 16
spinor ground state with creation operators $c_{-r}$, $r\in\ze+{1\over
2}$, tensored with states built up by creation oscillators $d_t$,
$t\in\ze\cup(\ze+{1\over 2})$, from momentum states $|\lambda\rangle$,
$\lambda\in E_8$. Thus, we have two pieces to the Virasoro operator on
the second factor, a piece constructed from the integer graded
oscillators and a piece disjoint from the rest. Both give $c=8$ and so
sum to give $c=16$, as required.

This is precisely the structure we would obtain by the approach of
\cite{Mythesis}, which gives us a form for the vertex operators. Note
that these are incorrect due to a normal ordering problem,
though they may be corrected by the argument used
in \cite{thesis}, which also specifies the procedure for obtain the
intertwining operators and hence the full orbifold theory.

It may seem that the application of the above techniques was
unnecessary, since we knew the explicit structure in any case. We have
merely used it as a simple example to demonstrate the power of our
approach, and in the next section will consider situations where no
similar argument can be applied. Note that we may turn the above
argument around, if desired, to show that the orbifold theory which we
may postulate explicitly in fact leads to a self-dual partition
function, as a consequence of the identity between \reg{FYA} and
\reg{twispart}.
\subsection{Application to the theories $\widetilde\Hil(\Lambda)$}
As we remarked above, the application to theories $\Hil(\Lambda)$ is
in some sense trivial, while the attempt at an analogue of the ternary
code $\ze$-lattice constructions, for which we have already stated that we
developed this approach, is work which is still in progress.
Let us instead consider projecting out by a
lattice induced involution the theories $\widetilde\Hil(\Lambda)$ for
$\Lambda$ self-dual. There are two cases to consider.

The first is that in which
$\widetilde\Hil(\Lambda)\cong\Hil(\Lambda')$  for some even
self-dual lattice $\Lambda'$
(of which there are 9
cases for $c=24$, {\em i.e.} one for each doubly-even self-dual binary linear
code in 24 dimensions\cite{DGMtrialsumm}).
In this case, it may be that the lattice induced
automorphism is not lattice induced from the point of view of the
theory $\Hil(\widetilde\Lambda)$, and so the above arguments would yield
non-trivial information about a new orbifold theory. However, it seems
likely that all the automorphisms of the theories $\Hil(\Lambda)$ are
given by the (extended) lattice automorphism group, and so this case is
probably not too promising, though the full automorphism group of
$\Hil(\Lambda)$ does still remain to be determined definitively.

The second possibility is that $\widetilde\Hil(\Lambda)\not\cong
\Hil(\Lambda')$ for any $\Lambda'$ (15 instances in $c=24$).
This means that we are certainly in a non-trivial situation, {\em
i.e.} the explicit twisted vertex operator construction of \cite{Mythesis} is
not valid.

A group of automorphisms of the theory (though not necessarily the full
automorphism group) $\widetilde\Hil(\Lambda)$ is
given by the exact sequence
\begin{equation}
1\rightarrow\Gamma(\Lambda)\rightarrow C(\Lambda)\rightarrow{\rm
Aut}\,(\Lambda)/\ze_2\rightarrow 1\,,
\end{equation}
where $\Gamma(\Lambda)=\{\pm\gamma_\lambda :\lambda\in\Lambda\}$, {\em
i.e.}
\begin{eqnarray}
u_{R,S}a^j_n{u_{R,S}}^{-1}&=R_{ij}a^i_n\,;\quad
u_{R,S}c^j_r{u_{R,S}}^{-1}&=R_{ij}c^i_r\\
u_{R,S}|\lambda\rangle&=v_{R,S}(\lambda)|R\lambda\rangle\,;\quad
u_{R,S}\chi&=S\chi\,,
\end{eqnarray}
where $R\in{\rm Aut}\,(\Lambda)$ and $S\gamma_\lambda
S^{-1}=v_{R,S}(\lambda)\gamma_{R\lambda}$. (See \cite{DGMtwisted} for a full
explanation of the notation. The $a_n$'s are the usual integer graded
oscillators
in the untwisted sector of $\Hil(\Lambda)$ acting on the momentum
ground states $|\lambda\rangle$,
while the $c_r$'s are half-integer graded oscillators in the twisted sector
acting on
the spinor ground states $\chi$, which form a representation space for the
gamma matrix algebra $\gamma_\lambda\gamma_\mu=(-1)^{\lambda\cdot\mu}
\gamma_\mu\gamma_\lambda$.)
Note that we only need to know the matrix $S$ for the evaluation of
$\alpha$, since the twisted states first appear at level 2 [We have
seen in the above that we can usually guess the value of $\alpha$ in
any case from Schellekens' list, but here we should really calculate
it as we are unable to work out the vacuum energy (and degeneracy)
like we were able to do in the previous case (where we knew the
twisted vertex operators explicitly), and so it would provide a
check, though not necessarily an independent one.]
However, we do need $v_{R,S}(\lambda)$ for
evaluating the number of invariant weight one states, except in some
special cases which we shall consider below.

Let us consider an example which falls into the second possibility
mentioned above. We have an automorphism given
by 6 pairs on transpositions on the root system components of the
Niemeier lattice ${A_2}^{12}$. Studying the glue code for this
lattice\cite{ConSlo}, we find that this automorphism is of order 4.
So, in particular, it would be unsuitable for the construction of a
$\ze_2$-orbifold from the theory $\Hil({A_2}^{12})$.
However, in the case of $\widetilde\Hil({A_2}^{12})$, the automorphism $\theta$
becomes
of order
2, since it squares to the lattice reflection, which has trivial
action on the twisted theory. The number of invariant weight one
states is given by
\begin{equation}
{1\over 2}{\rm Tr}_{\Hil_1}\,(1+\theta)\,,
\end{equation}
where $\Hil_1$ is the space of states at level one. Noting that
$v(\lambda)=v(-\lambda)$ (since we have chosen the ``gauge'' such that
$\gamma_{\lambda}=\gamma_{-\lambda}$\cite{DGMtwisted}),
we find that this becomes
\begin{equation}
{1\over
4}\sum_{\lambda\in\Lambda(2)}\left(1+v(\lambda)(\langle\lambda|
\theta\lambda\rangle + \langle -\lambda|\theta\lambda\rangle)\right)\,.
\end{equation}
In this case, $\langle\pm\lambda|\theta\lambda\rangle=0$, and so we
obtain ${1\over 4}|\Lambda(2)|$ invariant weight one states, leading
to ${1\over 4}|\Lambda(2)|+24(1-\alpha)$ weight one states in the
orbifold theory. Since ${1\over 4}|\Lambda(2)|=18$, the only possible
value for $\alpha$ (assuming the orbifold theory is consistent!) is 0.
Then the only possibility for the algebra is the new theory
${A_{2,4}}^4C_{2,4}$.
We know the invariant weight one states are of the form
$|\lambda\rangle + |-\lambda\rangle + |\theta\lambda\rangle +
|-\theta\lambda\rangle$. We can easily work out the corresponding algebra.
In the original twisted theory, the ${A_2}^{12}$ breaks down to
${A_1}^{12}$, and it is trivial to observe that the invariant algebra
in the new orbifold theory must be ${A_1}^6$.
The partition function argument tells us that 24 states arise from the
twisted sector to
enhance this algebra in the final orbifold theory.
The algebra ${A_{2,4}}^4C_{2,4}$ is thus at least consistent, in that
it contains the invariant algebra.

Consider the theory
$\widetilde\Hil({A_4}^6)$. This is also an example of the second case
referred to above, since it has algebra
${C_2}^6$. We have that set of pairwise transpositions on the 6
components of the root lattice exists in the automorphism group of the
lattice. (This exists since the glue code for ${A_4}^6$
is such that the group $G_2({A_4}^6)$ is isomorphic to $PGL_2(5)$ acting
on $\{\infty,0,1,2,3,4\}$\cite{ConSlo}. It is also an involution
acting on the twisted theory, as in the case of the above example.)
This automorphism will give us 54-24$\alpha$
weight one states in the new orbifold theory (assuming it is
consistent).
There are no theories
in Schellekens' list with $N=6$ or 54, and so we must have $\alpha=1$,
$N=30$, {\em i.e.} we have no algebra enhancement. Thus the algebra
must be just the invariant subalgebra, which is clearly
${C_{2,12}}^3$, an algebra which does appear on the list of section
\ref{Schell}.
This provides strong evidence of the existence of another new
theory.

Let us now make an attempt at a more systematic consideration.
One possibility would be to consider all automorphisms
given by transpositions of the components of the root system of the
Niemeier lattice
on the 15 non-trivial twisted lattice theories. For these, there is
no need to know the matrix $S$, except to confirm the value of $\alpha$, which
typically will be uniquely determined from Schellekens' list in any
case.
In general, in order to avoid consideration of the $v(\lambda)$
in the evaluation of the number of invariant weight one states, we
require $\theta\lambda$ to be distinct from both $\lambda$ and
$-\lambda$ for the vectors of length squared two. In particular,
distinction from $\lambda$ is equivalent to saying that, if the
automorphism does not interchange lattice components, it acts as a
no-fixed-point automorphism
(NFPA) on the root lattice of one component. (Conversely,
an NFPA on $\Lambda(2)$, the vectors of length squared 2 in $\Lambda$,
is also one on $\langle\Lambda(2)\rangle$.)
Distinction from $-\lambda$ means
that we take all NFPA's modulo the NFPA $-1$. The NFPA's of the root
lattices have been classified in \cite{Mythesis}.
The number of weight one states in the new orbifold theory is ${1\over
4}|\Lambda(2)|+24(1-\alpha)=6h+24(1-\alpha)$. This must be
greater than the number of invariant weight one states, which is
${1\over 4}|\Lambda(2)|$, and so we must have $\alpha=0$ or 1 if the
theory is to be consistent. In other words, consistency of the
orbifold theory requires that Moonshine holds!

Now, \cite{Mythesis} tells us that the number of
non-reflection NFP involutions of the ADE algebras is zero.
Let us therefore consider the 15 lattices giving rise to non-trivial
reflection twisted theories and try to orbifold with respect to an
NFPA of order 2 induced by transpositions and NFPA's of order 4 on the
root systems of the components. We summarise the results in table
\ref{invol}.

 From \cite{Mythesis}, we have that all automorphisms of the ADE root
systems of the same order are conjugate. Thus, all NFPA's of order 4
square to give the reflection involution, and so act as involutions on
the twisted theory, as required. The only root systems of type ADE
admitting fourth order NFPA's are $E_8$, $A_1$ and $D_{2n}$ for $2\leq
n\leq 11$ except for $n=9$\cite{Mythesis}. This, coupled with the
simple observation that many of the 15 theories cannot have
transposition automorphisms, allows us to immediately exclude a large
number of no-fixed-point involutions, indicated by a $\times$ in the
``Involution'' column of the table. The $\circ$ by the
Leech lattice just indicates that, since
it cannot give any interesting theories, we are not concerned with this
case. (The theories it produces must have either 24 or 0 weight one
states. The arguments of section \ref{uniqueness} and \cite{thesis}
then identify the
theories uniquely in this case, as we have indicated in the
appropriate columns of the table.)

As noted above,
the orbifold theory will have a number of weight one states given by
${1\over 4}|\Lambda(2)|+24-24\alpha$,
with $\alpha=0$ or $\alpha=1$. We list both possible numbers in the
table. Those for which no corresponding algebra exists on Schellekens'
list have a $\times$ by them. In those cases where there is a unique
algebra, we have noted that (in fact we know the unique theory in the
case of 24 weight one states\cite{DGMtriality} and have noted that).
The $\times$ by the algebra
${A_2}^4C_2$ in the ${A_6}^4$ entry indicates that, though this is
the only algebra in Schellekens' list with the appropriate value of
$N$, it cannot be the correct one since the involution must be a pure
transposition if it exists at all ({\em i.e.} if the automorphism
of the root system
extends to an automorphism of the glue code), and so we must have an
algebra of the form $X^2$ as the invariant algebra is not enhanced in
the case $\alpha=1$. The ? by the algebra ${C_2}^3$ in the ${A_4}^6$
entry indicates that, though there is no unique algebra for this
number of weight one states, this is the only one of the form $X^3$,
which we know the non-enhanced invariant algebra must be.

Referring to the
glue codes\cite{ConSlo} for the theories which we have not yet excluded from
having
involutions, we get transposition involutions as
indicated. Two of the cases marked by a ? in the ``Involution'' column
could have involutions given by
combining transpositions with fourth order root system automorphisms.
It remains to check both the glue code and the action of the fourth
order map on the corresponding glue vectors. The remaining ?
indicates that we have yet to investigate the corresponding glue code
for transposition automorphisms, though we see from the other columns
for this entry in the table that the orbifold cannot be consistent

The ${E_6}^4$ invariant algebra under $(1 -1)(1 -1)$ is ${C_4}^2$.
There is no such theory at $N=72$. We may ask if it can be enhanced
to one of the $N=96$
theories. We first note that we need at least 2 rank $\geq 4$ algebras
or one of rank $\geq 8$. This narrows down the list. Then a simple
consideration
of dimensions of the appropriate algebras shows that $C_4$ cannot be
embedded, except possibly as ${C_4}^2$ in $A_8$ to give the theory
${A_2}^2A_8$. This embedding though is clearly impossible, and so the
orbifold must be inconsistent.

The invariant algebra for ${A_7}^2{D_5}^2$ under $(1 -1)(1 -1)$ is
${C_2}^2D_4$.
Such an algebra exists at $N=48$. We can eliminate the possibility of
enhancement
to a theory at $N=72$ by trivially considering the possibilities with
the restriction
of one component of rank at least 4 and total rank 12. The postulated
orbifold theory is indicated in the table.

\begin{table}[htb]
\begin{center}
\begin{tabular}{|c|c|c|c|}\hline
Lattice & Involution & Orbifold I & Orbifold II \\ \hline
$\Lambda_{24}$ & $\circ$ & 24 $\Hil(\Lambda_{24})$ & 0 $V^\natural$ \\
${A_3}^8$ & $(1\leftrightarrow -1)\ldots
(1\leftrightarrow -1)$ & 48 & 24 $\Hil(\Lambda_{24})$ \\
${A_7}^2{D_5}^2$ & $(1\leftrightarrow -1)(-1\leftrightarrow 1)$ & 72 $\times$ &
48 ${C_2}^4D_4$? \\
$A_{24}$ & $\times$ & 174 $\times$ & 150 $\times$ \\
$A_{17}E_7$ & $\times$ & 132 $A_8F_4$ & 108 ${B_4}^3$ \\
$A_{15}D_9$ & $\times$ & 120 & 96 \\
${A_{12}}^2$ & $(1\leftrightarrow -1)$ & 102 $\times$ & 78 $\times$ \\
$A_{11}D_7E_6$ & $\times$ & 96 & 72 \\
${A_9}^2D_6$ & ? & 84 & 60 \\
${A_8}^3$ & $\times$ & 78 $\times$ & 54 $\times$ \\
${A_6}^4$ & ? & 66 $\times$ & 42 ${A_2}^4C_2$ $\times$ \\
${A_5}^4D_4$ & ? & 60 & 36 \\
${A_4}^6$ & $(1\leftrightarrow -1)(1\leftrightarrow
-1)(1\leftrightarrow -1)$
& 54 $\times$ & 30 ${C_2}^3$? \\
${A_2}^{12}$ &  $(1\leftrightarrow -1)\ldots
(1\leftrightarrow -1)$ & 42 ${A_2}^4C_2$ & 18 $\times$ \\
${E_6}^4$ & $(1\leftrightarrow -1)(-1\leftrightarrow 1)$ & 96 & 72 \\
\hline
\end{tabular}
\caption{No-fixed-point involution orbifolds of the 15 non-trivial
reflection twisted lattice theories.}
\label{invol}
\end{center}
\end{table}

\section{Conclusions}
Starting from results for CFT's by Schellekens analogous to
some of those proved for the
Niemeier lattices by Venkov, we have shown firstly how to reformulate
the uniqueness problem for the ``Monster'' CFT. The notion of
complementary representations coupled with the restrictions on
Kac-Moody algebras in the theories derived by Schellekens then enabled
us, for the first time, to investigate with some degree of confidence
orbifold theories for which no explicit construction is known.

It remains to present a more systematic survey of the orbifolds of the
theories $\widetilde\Hil(\Lambda)$ for arbitrary involutions. This
will require explicit knowledge of the action of the involution on the
twisted sector ground states, as discussed in \cite{GNOS}.
Also, we need to understand the conditions under which the orbifolds
are consistent. For the theories $\Hil(\Lambda)$, it seemed to be
sufficient simply to calculate the twisted sector ground state energy,
though for the theories $\widetilde\Hil(\Lambda)$ we need to find
some condition which will
eliminate the ${A_{12}}^2$ theory in table \ref{invol} say
while preserving the ${A_4}^6$ theory.
We could investigate this by evaluating the first few terms
in the partition function of the invariant sector, since our
assumption of $\Gamma_0(2)$ invariance
assumes an orbifold-like behaviour. In order to do this, we must again
know
about the action of the automorphism on the twisted sector ground
state.

Finally, the obvious hope for the future is to complete the analogue
of Venkov's results for CFT's. Venkov showed that, for each of the
possible semi-simple algebras which his conditions selected, that there was
one and only one corresponding even self-dual lattice. His approach
was based on coding theory via the idea of glue codes. (Any relation
to the code structures investigated in \cite{DGMtrialsumm,PSMcodes,thesis} is
as yet not understood.) However, the analogue cannot be exact, for
Schellekens has demonstrated that for some of the algebras listed in
section \ref{Schell} there is no modular invariant combination of
Kac-Moody algebras, and hence there can be no consistent orbifold
theory. Nevertheless, as an extension to the result of section
\ref{uniqueness}, it might be argued that, where a theory does exist,
it is unique. There are, as yet, no counterexamples to such a claim.

\section{Note}
After completion of this paper, the paper \cite{Schellekens2} was
received, in which the number of possible combinations of Kac-Moody
algebras listed in section \ref{Schell} was reduced to 71 by
consideration of higher order trace identities. The results would
appear to eliminate some of the new theories claimed in this paper,
although further checks remain to be done. Nevertheless, the
techniques discussed in this paper when applied to the remaining
involutions of the twisted lattice theories should be rendered more
powerful as a result of this more restrictive set of possibilities.
This is work which is currently in progress.

\end{document}